\RequirePackage{fix-cm}
\documentclass[twocolumn, epjc3, comma, sort&compress, natbib]{svjour3}
\bibpunct{[}{]}{,}{n}{}{,} 

\journalname{Eur. Phys. J. A}

\usepackage{latexsym}
\usepackage{amsmath}
\usepackage{amssymb}
\usepackage{amsfonts}
\usepackage{CJKutf8}

\usepackage[mathscr,scaled=1.15]{urwchancal}
\DeclareFontFamily{OT1}{pzc}{}
\DeclareFontShape{OT1}{pzc}{m}{it}%
{<-> s * [1.15] pzcmi7t}{}
\DeclareMathAlphabet{\mathpzc}{OT1}{pzc}{m}{it}

\usepackage{color}

\usepackage{supertabular}
\usepackage{placeins}
\usepackage{epsfig}
\usepackage{graphicx}

\definecolor{purple}{rgb}{0.5,0,0.5}
\definecolor{blue}{rgb}{0.0,0,0.9}
\definecolor{prdblue}{rgb}{0.133,0.118,0.498}
\usepackage[colorlinks=true, pdfstartview=FitV, linkcolor=prdblue, citecolor= prdblue, urlcolor=prdblue]{hyperref}

\makeatletter

\setbox0\hbox{$\xdef\scriptratio{\strip@pt\dimexpr
    \numexpr(\sf@size*65536)/\f@size sp}$}

\newcommand{\scriptveryshortarrow}[1][3pt]{{%
    \hbox{\rule[\scriptratio\dimexpr\fontdimen22\textfont2-.2pt\relax]
               {\scriptratio\dimexpr#1\relax}{\scriptratio\dimexpr.4pt\relax}}%
   \mkern-4mu\hbox{\let\f@size\sf@size\usefont{U}{lasy}{m}{n}\symbol{41}}}}

\makeatother

\begin{document}

\begin{CJK}{UTF8}{song}

\title{$\,$\\[-6ex]\hspace*{\fill}{\normalsize{\sf\emph{Preprint no}. NJU-INP 065/22}}\\[1ex]
Bethe-Salpeter kernel and properties of strange-quark mesons}

\author{Zhen-Ni Xu\thanksref{NJU,INP}
       $\,^{\href{https://orcid.org/0000-0002-9104-9680}{\textcolor[rgb]{0.00,1.00,0.00}{\sf ID}}}$
    \and
        Zhao-Qian Yao\thanksref{NJU,INP,ECT}
       $\,^{\href{https://orcid.org/0000-0002-9621-6994}{\textcolor[rgb]{0.00,1.00,0.00}{\sf ID}}}$
    \and
        \\Si-Xue Qin\thanksref{CQU}
       $\,^{\href{https://orcid.org/0000-0002-6754-6046}{\textcolor[rgb]{0.00,1.00,0.00}{\sf ID}}}$
    \and
        Zhu-Fang Cui\thanksref{NJU,INP}
       $\,^{\href{https://orcid.org/0000-0003-3890-0242}{\textcolor[rgb]{0.00,1.00,0.00}{\sf ID}}}$
    \and
        Craig D.~Roberts%
       $\,^{\href{https://orcid.org/0000-0002-2937-1361}{\textcolor[rgb]{0.00,1.00,0.00}{\sf ID}}}$
}

\authorrunning{Z.-F.~Cui \emph{et al}.} 

\institute{School of Physics, Nanjing University, Nanjing, Jiangsu 210093, China \label{NJU}
           \and
           Institute for Nonperturbative Physics, Nanjing University, Nanjing, Jiangsu 210093, China \label{INP}
           \and
           European Centre for Theoretical Studies in Nuclear Physics
            and Related Areas, \\\hspace*{1ex}Villa Tambosi, Strada delle Tabarelle 286, I-38123 Villazzano (TN), Italy \label{ECT}
           \and
            Department of Physics, Chongqing University, Chongqing 401331, China \label{CQU}
\\[1ex]
Email:
\href{mailto:sqin@cqu.edu.cn}{sqin@cqu.edu.cn} (S.-X. Qin);
\href{mailto:phycui@nju.edu.cn}{phycui@nju.edu.cn} (Z.-F. Cui);
\href{mailto:cdroberts@nju.edu.cn}{cdroberts@nju.edu.cn} (C. D. Roberts)
            }

\date{2023 February 08}

\maketitle

\end{CJK}

\begin{abstract}
Focusing on the continuum meson bound-state problem, a novel method is used to calculate closed-form Bethe-Salpeter kernels that are symmetry consistent with any reasonable gluon-quark vertex, $\Gamma_\nu$, and therewith deliver a Poincar\'e-invariant treatment of the spectrum and decay constants of the ground- and first-excited states of $u$, $d$, $s$ mesons.  The predictions include masses of as-yet unseen states and many unmeasured decay constants.  The analysis reveals that a realistic, unified description of meson properties (including level orderings and mass splittings) requires a sound expression of emergent hadron mass in bound-state kernels; alternatively, that such properties may reveal much about the emergence of mass in the standard model.
\end{abstract}



%
\section{Introduction}
Notwithstanding fifty years of quantum chromodynamics (QCD), the spectrum of light quark mesons remains obscure.  This is especially true if one includes states with strangeness, \emph{e.g}., Refs.\,\cite{Ketzer:2019wmd} and \cite[Sec.\,63]{Workman:2022ynf}, and despite the claim, made long ago \cite{Godfrey:1985xj}, ``\ldots that all mesons -- from the pion to the upsilon -- can be described in a unified framework.''  The latter statement was made with reference to a quark model potential with one-gluon-like exchange plus linear confinement.  Some features of such potential models can be understood to have a qualitative connection with QCD.  For instance, regarding the lighter degrees-of-freedom, $g=u$, $d$, $s$ quarks, one might draw a link between a model's constituent quarks and QCD's dressed-quarks, insofar as dressed quarks are described by a momentum-dependent running-mass, $M_g(k^2)$, which is large at infrared momenta \cite[Fig.\,2.5]{Roberts:2021nhw}: $M_{u,d}(0)\simeq 0.41\,$GeV, $M_s(0) \simeq 0.53\,$GeV.

There are problems with the quark model position, however.
One issue is the quantum mechanics description of light-quark systems in terms of a potential.  Owing to the fact that light-particle annihilation and creation effects are essentially nonperturbative in QCD, it has thus for proved impossible to calculate a quantum mechanical potential between two light quarks \cite{Bali:2005fu, Prkacin:2005dc, Chang:2009ae}.
Another recognises that whilst it might be possible to connect a linear potential with the Wilson area law applicable to infinitely-heavy colour-sources and -sinks \cite{Wilson:1974sk}, the associated flux tube picture \cite{Isgur:1983wj} has neither a mathematical nor physical connection with the confinement of light quarks.  In this sector, at least, confinement is more subtle, arguably manifested in analytic properties of dressed-gluon and \mbox{-quark} propagators that are markedly different from those of asymptotic states
\cite{Krein:1990sf, Burden:1991gd, Brodsky:2012ku, Qin:2013ufa, Lucha:2016vte, Gao:2017uox, Binosi:2019ecz, Dudal:2019gvn, Fischer:2020xnb, Roberts:2020hiw, Ding:2022ows}.
A third notes that even if these and other problems could be overcome, there is the Gell-Mann--Oakes--Renner (GMOR) identity \cite{GellMann:1968rz}, which linearly relates the pion mass-squared, $m_\pi^2$, to Nature's explicit source of chiral symmetry breaking, $\hat m$, \emph{viz}.\ $m_\pi^2 \propto \hat m$.  Such behaviour is impossible in a potential model \cite{Horn:2016rip}.

Challenges like that can be surmounted by using continuum Schwinger function methods (CSMs) \cite{Eichmann:2016yit, Qin:2020rad}, \emph{e.g}., the GMOR identity and it corollaries are naturally obtained \cite{Maris:1997tm, Qin:2014vya, Holl:2005vu}.
Yet, other issues arise.  The most notable relates to construction of a practicable, symmetry preserving approximation to the quark+antiquark Bethe-Salpeter kernel, where the symmetries include vector current conservation, partial conservation of the axialvector current, Poincar\'e-covariance, etc.  A straightforward systematic approach \cite{Munczek:1994zz, Bender:1996bb} serves well for ground-state hadrons with little rest-frame orbital angular momentum between the dressed valence consti\-tuents \cite{Holl:2005vu, Fischer:2009jm, Krassnigg:2009zh, Qin:2011dd, Qin:2011xq, Blank:2011ha, Hilger:2014nma, Fischer:2014cfa, Eichmann:2016hgl, Qin:2019hgk}.  It is limited, however, by a failure to express consequences of emergent hadron mass (EHM) \cite{Roberts:2020udq, Roberts:2020hiw, Roberts:2021xnz, Roberts:2021nhw, Binosi:2022djx, Papavassiliou:2022wrb, Ding:2022ows, Roberts:2022rxm, Ferreira:2023fva, Carman:2023zke}.  Improved schemes are being developed \cite{Chang:2009zb, Chang:2011ei, Binosi:2014aea, Williams:2015cvx, Binosi:2016rxz, Qin:2020jig}; and in applications to mesons constituted from $u$, $d$ valence quarks and/or antiquarks, to which they have hitherto been restricted, the new approaches have shown promise.

%

\begin{figure*}[!t]
\centerline{%
\includegraphics[clip, width=0.9\textwidth]{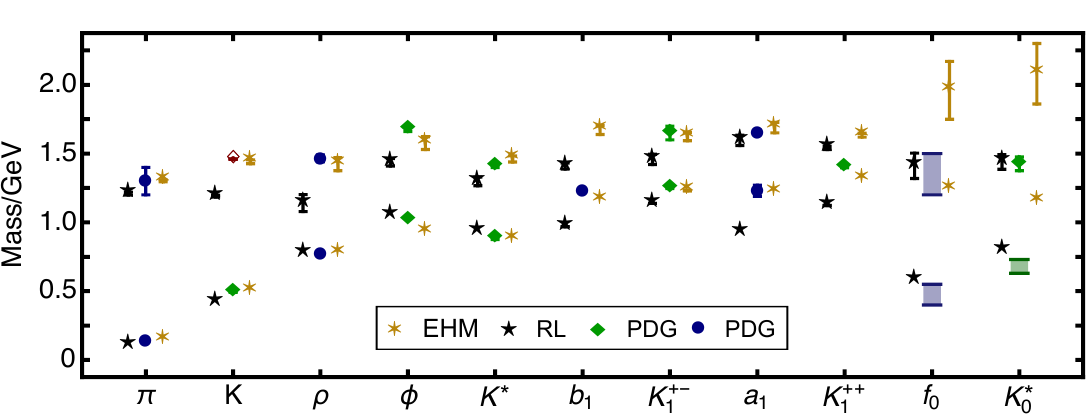}}
\caption{\label{figAllstrange}
Gold six-pointed stars -- spectrum of low-lying $u$, $d$, $s$ mesons predicted by the Bethe-Salpeter kernel developed herein;
and black five-pointed stars -- same spectrum computed using rainbow-ladder truncation.
Comparison empirical spectrum \cite[Summary Tables]{Workman:2022ynf}:
blue circles (bars) -- $u$, $d$ systems; and green diamonds (bar) -- mesons with $s$ and/or $\bar s$ quarks.  The open red diamond is the $K(1460)$, about which little is known.
%
}
\end{figure*}

Herein, we explore the capacity of a new kernel construction \cite{Qin:2020jig} to simultaneously treat ground- and first-excited-states of light-quark mesons plus those containing $s$ and/or $\bar s$ quarks.  As shown in Fig.\,\ref{figAllstrange}, the empirical spectrum displays some curious features.  For instance:
whilst $m_\rho < m_{K^\ast}$, this ordering is reversed for the first excitations of these states;
the first excited state of the $\rho$ is heavier than that of the $\pi$, but this is reversed for $K^\ast$, $K$;
and there is a near degeneracy between axialvector mesons, with the heavier mass of the $s$ quarks seeming to have little or no impact.
In delivering the first Poincar\'e covariant, symmetry-preserving analysis of this collection of mesons to employ an EHM-improved kernel, we provide fresh insights into the nature of lighter-quark mesons.

\section{Rainbow-ladder truncation}
Mesons appear as poles in the quark+antiquark scattering matrix: the pole location reveals the mass (and width); and the residue of the pole is the Poincar\'e-covariant bound-state wave function.  The scattering matrix is the solution of an integral equation, whose driving term is the Bethe-Salpeter kernel and which features the dressed-propagators of the valence degrees-of-freedom that constitute the system.

The simplest symmetry-preserving approximation to the meson problem is provided by the rainbow-ladder (RL) truncation, which is leading-order in the scheme identified in Refs.\,\cite{Munczek:1994zz, Bender:1996bb}.  It may be introduced by focusing on the gap equation for a quark with flavour $g$:
\begin{subequations}
\label{EqGap}
\begin{align}
S_g^{-1}(k) &= i\gamma\cdot k + m_g + \Sigma^g(k) \,,\\
\Sigma^g(k) & =\int_{dq}
4 \pi \alpha\, D_{\mu\nu}(l)\gamma_\mu\frac{\lambda^{a}}{2} S(q) \Gamma_\nu^g(q,k)\frac{\lambda^{a}}{2}\,,
\end{align}
\end{subequations}
where
$l=k-q$,
$m_g$ is the quark current-mass;
$\{\tfrac{1}{2}\lambda^a|a=1,\ldots,8\}$ are the generators of SU$(3)$-colour in the fundamental representation;
$\alpha$ is the QCD coupling; $D_{\mu\nu}$ is the dressed-gluon propagator; and $\Gamma_\nu^g$ is the relevant dressed-gluon-quark vertex.  The solution of Eq.\,\eqref{EqGap} is often written
$S_g(k) = 1/[i\gamma\cdot k\,A^g(k^2) + B^g(k^2)]$.

Following Refs.\,\cite{Munczek:1994zz, Bender:1996bb}, the Bethe-Salpeter kernel is determined once the diagrammatic content of the kernel in Eq.\,\eqref{EqGap} is specified.  The rainbow-truncation is obtained by writing:
$4 \pi \alpha D_{\mu\nu}(l) \Gamma_\nu^g(q,k) \to {\mathpzc G}_{\mu\nu}(l)\gamma_\nu$,
where ${\mathpzc G}_{\mu\nu}$ is a vector-boson exchange-interaction informed by analyses of QCD's gauge sector \cite{Qin:2011dd, Qin:2011xq, Binosi:2014aea}.  In the associated ladder truncation, the Bethe-Salpeter kernel is $(y=l^2)$
\begin{subequations}
\label{EqRLInteraction}
\begin{align}
\label{KDinteraction}
\mathscr{K}_{tu}^{rs} & = {\mathpzc G}_{\mu\nu}(l) [i\gamma_\mu\frac{\lambda^{a}}{2} ]_{ts} [i\gamma_\nu\frac{\lambda^{a}}{2} ]_{ru}\,,\\
 {\mathpzc G}_{\mu\nu}(l)  & = \tilde{\mathpzc G}(y) T_{\mu\nu}(l)\,,
\end{align}
\end{subequations}
$l^2 T_{\mu\nu}(l) = l^2 \delta_{\mu\nu} - l_\mu l_\nu$.  This tensor structure specifies Landau gauge, which is used because \cite{Bashir:2009fv}: (\emph{i}) it is a fixed point of the renormalisation group; (\emph{ii}) that gauge for which corrections to RL truncation are least noticeable; and (\emph{iii}) most readily implemented in lattice-regularised QCD.
In Eq.\,\eqref{EqRLInteraction}, $r,s,t,u$ represent colour, spinor, and flavour matrix indices (as necessary).

A useful form of ${\mathpzc G}_{\mu\nu}(l)$ is explained in Refs.\,\cite{Qin:2011xq, Binosi:2014aea}:
%
\begin{align}
\label{defcalG}
 \tilde{\mathpzc G}(y) & =
 \frac{8\pi^2 D}{\omega^4} e^{-y/\omega^2} + \frac{8\pi^2 \gamma_m \mathcal{F}(y)}{\ln\big[ \tau+(1+y/\Lambda_{\rm QCD}^2)^2 \big]}\,,
\end{align}
where $\gamma_m=4/\beta_0$, $\beta_0=25/3$,
$\Lambda_{\rm QCD}=0.234\,$GeV,
$\ln(\tau+1)=2$,
and ${\cal F}(y) = \{1 - \exp(-y/[4 m_t^2])\}/y$, $m_t=0.5\,$GeV.
Here, we only recapitulate a few details so as to provide context for a simplification that we subsequently implement.
%
(\emph{i}) $0 < \tilde{\mathpzc G}(0) < \infty$ because a nonzero gluon mass-scale appears as a consequence of EHM in QCD \cite{Binosi:2014aea, Cui:2019dwv, Roberts:2021nhw, Binosi:2022djx, Papavassiliou:2022wrb, Ding:2022ows};
and (\emph{ii}) the large-$y$ behaviour ensures that the one-loop renormalisation group flow of QCD is preserved.
Property (\emph{ii}) is crucial when considering, \emph{e.g}., hadron elastic and transition form factors at large momentum transfer \cite{Chen:2018rwz, Ding:2018xwy, Xu:2019ilh, Xu:2021mju} and the behaviour of parton distribution functions and amplitudes near the endpoints of their support domains \cite{Ding:2019qlr, Cui:2020tdf}.
Notably, it is far less important when calculating masses, which are global, integrated properties.

Regarding masses, (\emph{i}) is critical: even a symmetry-preserving treatment of a momentum-independent interaction can deliver good results \cite{Yin:2021uom, Xu:2021iwv, Gutierrez-Guerrero:2021rsx}.  Hence, we follow Refs.\,\cite{Chang:2009zb, Chang:2010hb, Qin:2020jig}, hereafter retaining only the first term on the right-hand-side of Eq.\,\eqref{defcalG}.  By obviating the need for renormalisation, this simplifies analysis without materially affecting results.

\begin{table}[th]
\caption{\label{TableMassPredictions}
Masses calculated using RL and EHM-improved kernels compared with central values reported in Ref.\,\cite[PDG\,--\,Summary Tables]{Workman:2022ynf}: $n=0$, ground states; and $n=1$ first excited states.
(All results in GeV.  Where relevant, calculated results show Pad\'e-fit extrapolation uncertainty.)
 }
\begin{center}
\begin{tabular*}
{\hsize}
{
l@{\extracolsep{0ptplus1fil}}
l@{\extracolsep{0ptplus1fil}}
|l@{\extracolsep{0ptplus1fil}}
l@{\extracolsep{0ptplus1fil}}
l@{\extracolsep{0ptplus1fil}}}
& & RL & PDG & EHM-improved \\\hline
$\pi$ & $n=0\ $ & $0.103\ $ & $0.138\ $ & $0.140\ $ \\
 & $n=1\ $ & $1.209(11)\ $ & $1.3(1)\ $ & $1.302(7)\ $ \\\hline
$K$ & $n=0\ $ & $0.415\ $ & $0.494\ $ & $0.494\ $ \\
 & $n=1\ $ & $1.191(7)\ $ & $1.460\ $ & $1.440(14)\ $ \\\hline
$\rho$ & $n=0\ $ & $0.77\ $ & $0.775\ $ & $0.77\ $ \\
 & $n=1\ $ & $1.140(62)\ $ & $1.465(25)\ $ & $1.423(48)\ $ \\\hline
$\phi$ & $n=0\ $ & $1.049\ $ & $1.020\ $ & $0.926\ $ \\
 & $n=1\ $ & $1.437(27)\ $ & $1.680(20)\ $ & $1.576(46)\ $ \\\hline
$K^\ast$ & $n=0\ $ & $0.936\ $ & $0.890(14)\ $ & $0.872\ $ \\
 & $n=1\ $ & $1.298(32)\ $ & $1.414(15)\ $ & $1.461(22)\ $ \\\hline
$b_1$ & $n=0\ $ & $0.968\ $ & $1.230(3)\ $ & $1.159\ $ \\
& $n=1\ $ & $1.409(22)\ $ &  & $1.671(32)\ $ \\\hline
$K_1^{+-}$ & $n=0\ $ & $1.138\ $ & $1.253(7)\ $ & $1.230\ $ \\
& $n=1\ $ & $1.458(37)\ $ & $1.650(50)\ $ & $1.623(30)\ $ \\\hline
$a_1$ & $n=0\ $ & $0.929\ $ & $1.230(40)\ $ & $1.218\ $ \\
& $n=1\ $ & $1.596(37)\ $ & $1.655(16)\ $ & $1.689(38)\ $ \\\hline
$K_1^{++}$ & $n=0\ $ & $1.123\ $ & $1.403(7)\ $ & $1.309\ $ \\
& $n=1\ $ & $1.541(12)\ $ &  & $1.634(14)\ $ \\\hline
$f_0$ & $n=0\ $ & $0.577\ $ & $0.4-0.55\ $ & $1.237\ $ \\
& $n=1\ $ & $1.411(92)\ $ & $1.2-1.5\ $ & $1.96(21)\ $ \\\hline
$K_0^\ast$ & $n=0\ $ & $0.794\ $ & $0.63-0.73\ $ & $1.154\ $ \\
& $n=1\ $ & $1.439(53)\ $ & $1.425(50)\ $ & $2.08(22)\ $ \\\hline
\end{tabular*}
\end{center}
\end{table}

A typical RL truncation result for the spectrum of $u$, $d$, $s$ mesons is depicted in Fig.\,\ref{figAllstrange}, which provides a readily assimilated visual rendering of the results listed in Table~\ref{TableMassPredictions}.
The spectrum was obtained by solving the coupled gap and Bethe-Salpeter equations with the interaction just described, using $\omega = 0.8\,$GeV, a value chosen because it matches results from analyses of QCD's gauge sector \cite{Binosi:2014aea, Cui:2019dwv}, $D=D_{\rm RL}$, with
\begin{equation}
\label{DRL}
 \omega D_{\rm RL} = (1.01\,{\rm GeV})^3\,,
\end{equation}
and current-quark masses
$m_u = m_d = 2.7\,$MeV, $m_s=72\,$MeV.
Since we have chosen to work with an interaction that eliminates the need for renormalisation, these current masses need not match those inferred from experiment.  Nevertheless, compared with such values \cite{Workman:2022ynf}: our mean $u$, $d$ current mass is not markedly different, \emph{cf}.\ $3.45_{-0.15}^{+0.35}\,$MeV; and our value for the ratio $2 m_s/(m_u+m_d) = 26.7$ is a good match, \emph{cf}.\ $27.3_{-0.8}^{+0.7}$.

The features and flaws of RL truncation are evident in Fig.\,\ref{figAllstrange}.  The mean absolute relative difference between RL masses and central experimental values is $13(8)$\%.  Whilst this might appear to be fair agreement, there is substantial scatter; and there are many qualitative discrepancies.  For instance, labelling the first excited state with an apostrophe, then:
$m_{K^\prime} < m_{\pi^\prime}$ in RL truncation, whereas the empirical ordering is opposite, and
the same is true for $(m_{\rho^\prime},m_{\pi^\prime})$,  $(m_{\rho^\prime},m_{K^{\ast \prime}})$;
RL truncation $a_1$-$\rho$ and $b_1$-$\rho$ mass splittings are one-third of the empirical values because the $b_1$ and $a_1$ mesons are much too light;
$m_{\phi^\prime}-m_\phi$ is half the experimental value;
and the level ordering of the $K_1^{+-}$, $K_1^{++}$ states is incorrect.

Furthermore, RL truncation produces light \linebreak quark+antiquark scalar mesons, which are not seen in Nature.  As explained elsewhere \cite[Sec.\,64]{Workman:2022ynf}, the lightest scalar mesons are now considered to be complicated systems with material meson+antimeson components.  Thus, the apparent agreement between experiment and the masses of the purely quark+antiquark $f_0$ and $K_0^\ast$ mesons generated by RL truncation is misleading.  Viewed from our perspective, the kernels described herein generate a hadron's dressed-quark core.  They do not include the resonant contributions which are typically associated with a meson-cloud.  Hence, the quark-core masses of any purely quark+antiquark $f_0$ and $K_0^\ast$ mesons should be significantly greater than the empirical value because adding resonant contributions to the bound-state kernels will generate a large amount of attraction.  This is illustrated, \emph{e.g}., in Refs.\,\cite{Holl:2005st, Santowsky:2020pwd}.

Some states listed in Ref.\,\cite[PDG\,--\,Summary Tables]{Workman:2022ynf} are omitted from our study for the reasons enumerated here.
(\emph{a}) Isospin partners are not distinguished by existing Bethe-Salpeter kernels; so, \emph{e.g}., the $\rho$- and $\omega$-mesons are degenerate.  Meson+meson final state interactions are necessary to remove such degeneracies, as highlighted elsewhere \cite{Pichowsky:1999mu, Chen:2017mug}.
(\emph{b}) Any realistic analysis of the $\eta$-$\eta^\prime$ complex must employ a kernel that expresses effects generated by the non-Abelian anomaly \cite{Christos:1984tu, Horvatic:2018ztu, Bhagwat:2007ha}.  This can be argued to modify Eq.\,\eqref{eqAVWTI} in the manner explicated in Refs.\,\cite{Ding:2018xwy}.  We will seek a unification of that approach and ours elsewhere.
(\emph{c}) Given that the scalar mesons $f_0(980)$, $a_0(980)$ are isospin partners whose properties are much affected by meson+meson final state interactions \cite[Sec.\,64]{Workman:2022ynf}, we have nothing to contribute herein beyond that which has already been noted in connection with the $f_0(500)$ and $K_0^\ast(700)$ states.

It is also worth commenting on our identification of the $K_1^{+-}$ and $K_1^{++}$ states, which is unrelated to quark model notions.  We solve the Bethe-Salpeter equation in the $J^P=1^+$ channel.  For both $n=0, 1$, there are two nigh-degenerate but distinct solutions; and since $m_s \neq m_{u,d}$, neither is $J^{PC}=1^{++}$.
On the other hand, considering $n=0$, the Bethe-Salpeter amplitude for one of the states is nearly antisymmetric under $k\cdot P \to -k\cdot P$ and its leptonic decay constant is very small -- see Table~\ref{TableDecayConstantPredictionsn0}.  These qualities identify this solution as a physical progeny of the $1^{+-}$ nonet mesons; hence, we label it $K_1^{+-}$.
For the other $n=0$ state, the Bethe-Salpeter amplitude is nearly symmetric under $k\cdot P \to -k\cdot P$ and its leptonic decay constant is much the same as that of the $a_1$ meson; consequently, it is most closely related to the $1^{++}$ meson nonet.
Similar features are manifest in the $n=1$ states.


A technical remark is also in order.
The Bethe-Salpe\-ter equation can figuratively be written as an eigenvalue problem, \emph{viz}.\ $\lambda(m) \Gamma(m) = K(m) \Gamma(m)$, where $K$ is the kernel and $\Gamma$ is the bound-state amplitude.  The on-shell solution is obtained at that mass $m$ for which the eigenvalue is unity, \emph{i.e}., $d(m):=1-\lambda(m)=0$.
Using RL truncation, one must employ an artificially inflated value of $D$ in Eq.\,\eqref{DRL} in order to approach a realistic expression of EHM \cite{Binosi:2014aea}.
Consequently, poles in the dressed-quark propagators enter the complex-plane integration domain sampled by the Bethe-Salpeter equation at lower values of $m$ than might otherwise be the case \cite{Maris:1997tm, Windisch:2016iud}.
This limits the number of states for which the Bethe-Salpeter equation can be solved directly using simple numerical algorithms.
In our case, this is the set of all meson ground-states in Fig.\,\ref{figAllstrange}. 
For each excited-state meson, we develop a $[n,l]$ Pad\'e approximant to $d(m)$, $1\leq n\leq 4$, $l\leq n$, and determine the zero by extrapolation.  Using a one-point jackknife procedure, we select the result obtained with those values of $n$, $l$ for which the uncertainty in the zero's location is smallest, reporting the associated uncertainty.

\section{EHM improved kernel}
\label{SecEHM}
The key to a realistic expression of EHM in mesons lies in using gap equation kernels with a more direct connection to QCD.  Today, $\alpha D_{\mu \nu}$ is well understood \cite{Binosi:2014aea, Cui:2019dwv, Roberts:2021nhw, Binosi:2022djx, Papavassiliou:2022wrb, Ding:2022ows}; and although much remains to be learnt about $\Gamma_\nu^g$, EHM is known to generate a large anomalous chromomagnetic moment (ACM) for the lighter quarks \cite{Chang:2010hb, Singh:1985sg, Bicudo:1998qb, Bashir:2011dp, Binosi:2016wcx, Kizilersu:2021jen} and, as illustrated elsewhere \cite{Chang:2011ei, Williams:2015cvx, Qin:2020jig}, this ACM has a marked impact on the $u$, $d$ meson spectrum.

To expand existing studies and expose novel ACM effects on mesons containing $s$ and/or $\bar s$ quarks, we write \cite{Qin:2020jig} ($\sigma_{l\nu} = \sigma_{\rho\nu} l_\rho$,
$\sigma_{\rho\nu}= (i/2)[\gamma_\rho,\gamma_\nu]$)
\begin{equation}
\label{EqVertexGap}
\Gamma_\nu^g(q,k)  = \gamma_\nu + \tau_\nu(l)\,,\; \tau_\nu(l)
= \eta \kappa(l^2)\sigma_{l\nu} \,,
\end{equation}
$\kappa(l^2) = (1/\omega)\exp{(-l^2/\omega^2)}$.
Here, $\tau_\nu(l)$ is the ACM term, with $\eta$ its strength.
In QCD, $\kappa(l^2)$ is power-law suppressed in the ultraviolet; but the Gaussian form, alike with the infrared-dominant term in Eq.\,\eqref{defcalG}, is sufficient for illustrative purposes.
Following RL truncation convention, any overall dressing factor $F_1$, as in $F_1(l^2)[\gamma_\nu + \tau_\nu(l)]$, is implicitly absorbed into $\tilde{\mathpzc G}(l^2)$.
Equation~\eqref{EqVertexGap} assumes the vertex is flavour-independent.  This is a good approximation for the lighter quarks \cite{Bhagwat:2004hn, Williams:2014iea}.

Having specified the gap equation kernel, without reference to its diagrammatic content, the method of Ref.\,\cite{Qin:2020jig} can be used to obtain a (continuous and discrete) symmetry-consistent closed-form for the Bethe-Salpeter kernel.  To proceed, consider the inhomogeneous Bethe-Salpeter equation, written figuratively \linebreak ($g,h=u,d,s$):
\begin{align}
\Gamma_{H\alpha\beta}^{gh}(k;P) & = {\mathpzc g}_H + \int_{dq}K^{(2)}_{\alpha\alpha^\prime,\beta^\prime\beta}
\chi_{H\alpha^\prime \beta^\prime}^{gh} (q,P)\,, \label{BSequation}
\end{align}
where $P=k_+-k_-$ is the total momentum of the quark($k_+)$+antiquark$(k_-)$ system;
${\mathpzc g}_H$ is a Dirac matrix combination that specifies the $J^{P(C)}$ of the channel under consideration;
$K^{(2)}$ is the two-particle irreducible quark+antiquark scattering kernel, carrying one index for each of the four fermion legs;
$\int_{dq}$ denotes a four dimensional integral;
and $\chi_H^{gh} (k,P) = S_f(k_+)\Gamma_H^{gh}(k,P)$ $S_g(k_-)$ is the unamputated vertex.  On an $H$-meson mass-shell, $\Gamma_H^{gh}$ is the bound-state amplitude, with $\chi_H^{gh}$ the associated Poincar\'e-covariant wave  function.

In systems that may include nondegenerate valence quarks, Ref.\,\cite[Eqs.\,(6)]{Qin:2020jig} must be generalised by considering the following forms for the Ward-Green-Takahashi (WGT) identities satisfied, respectively, by the unamputated inhomogeneous vector and axialvector vertices -- ${\mathpzc g}^H = i \gamma_\mu, \gamma_\mu\gamma_5$ in Eq.\,\eqref{BSequation}:
{\allowdisplaybreaks
\begin{subequations}
\label{eq:WTI}
\begin{align}
	P_\mu \chi_{\mu}^{gh}& (k_+,k_-) \nonumber \\
 & = i\Delta_{S_{gh}}^{\pm}(k) + i (m_g-m_h) \chi_{0}^{gh}(k_+,k_-) \,, \\
%
P_\mu \chi_{5\mu}^{gh}&(k_+,k_-) \nonumber \\
& =  i \Delta_{S5}^{\pm}(k) - i (m_g+m_h) \chi_{5}^{gh}(k_+,k_-)\,, \label{eqAVWTI}
\end{align}
\end{subequations}
where
$\Delta_{F_{gh}}^{\pm}(k) = F_g(k_+)-F_h(k_-)$,
$\Delta_{F5_{gh}}^{\pm}(k) = F_g(k_+) $ $\gamma_5+ \gamma_5 F_h(k_-)$,
and $\chi_{0/5}^{gh}$ are the unamputated \linebreak scalar/pseudoscalar vertices (${\mathpzc g}^H = {\mathbb I}/ \gamma_5$).
}


Subsequently repeating the algebra that leads from Eqs.\,(7) to (10) in Ref.\,\cite{Qin:2020jig}, one arrives at analogues of the four entries in Ref.\,\cite[Eq.\,(S.6)]{Qin:2020jig}, which are recorded in Eqs.\,\eqref{SuppEqs} herein.
Importantly, when $m_g \neq m_h$, the identities satisfied by $\Sigma^g_B(k_+)$, $\Sigma^h_B(k_-)$ are not related by charge conjugation; so, in contrast to \cite[Eqs.\,(10)]{Qin:2020jig}, herein one has four independent constraint equations.

Now introducing Eq.\,\eqref{EqVertexGap} into the gap equations, \linebreak Eq.\,\eqref{EqGap}, and using the algebraic equations obtained \linebreak thereby to construct explicit forms for the WGT identity constraints, Eqs.\,\eqref{SuppEqs}, one finds
\begin{align}
	{K}^{(2)} &= - \mathcal{G}_{\mu\nu}(l)\gamma_\mu\otimes\gamma_\nu  - \mathcal{G}_{\mu\nu}(l)\gamma_\mu \otimes \tau_\nu(l) \notag\\
	& + ~ \mathcal{G}_{\mu\nu}(l) \tau_\nu(l) \otimes \gamma_\mu +  {K}_{\rm ad} \,.
\end{align}
%
Equation\,\eqref{S.6c} is blind to ${K}_{\rm ad}$, but that is not the case for the other entries in the block.  So, writing
\begin{align}
	{K}_{\rm ad} &= [ \mathbf{1} \otimes_+ \mathbf{1} ] f^{(+)}_{p0} + [ -\mathcal{G}_{\mu\nu}(l)\gamma_\mu \otimes_+ \gamma_\nu ] f^{(-)}_{p1}  \notag\\
	& \quad + [ \mathbf{1} \otimes_- \mathbf{1} ] f^{(+)}_{n0} + [ -\mathcal{G}_{\mu\nu}(l)\sigma_{l\mu} \otimes_- \sigma_{l\nu} ] f^{(+)}_{n1}\,,
	\label{eq:kernel_Ad}
\end{align}
where $\otimes_\pm := \tfrac{1}{2}(\otimes \pm \gamma_5 \otimes \gamma_5)$ and
$f_{pj}^{(\pm)} = {\mathsf u}_{pj}^{(\pm)}(l^2,P^2) + i {\mathsf v}_{pj}^{(\pm)}(l^2,P^2)$,
$f_{nj}^{(+)} = {\mathsf u}_{nj}^{(+)}(l^2,P^2)$, $ j=0,1$, with ${\mathsf u}(l^2;P^2)$, ${\mathsf v}(l^2;P^2)\in \mathbb{R}$ for $\{l^2,P^2\}\in \mathbb{R}$,
one arrives at the following integral equations,
{\allowdisplaybreaks
\begin{subequations}
\label{KernelCompletion}
\begin{align}
& \int_{dq} \, {\mathpzc G}_{\mu\nu}(l) \gamma_\mu {\mathpzc s}_A^g(q_+) \tau_\nu(l) \nonumber\\
= & \int_{dq} \, [{\mathpzc s}_B^g(q_+) f_{p0}^{(+)} + {\mathpzc G}_{\mu\nu}(l)\gamma_\mu
{\mathpzc s}_B^h(q_-) \gamma_\nu f_{p1}^{(-)}]\,, \\
& \int_{dq} \, {\mathpzc G}_{\mu\nu}(l) \gamma_\mu {\mathpzc s}_B^g(q_+) \tau_\nu(l) \nonumber\\
= & \int_{dq} \, [{\mathpzc s}_A^g(q_+) f_{n0}^{(+)} - {\mathpzc G}_{\mu\nu}(l)\sigma_{l\mu} {\mathpzc s}_A^g(q_+) \sigma_{l\nu} f_{n1}^{(+)}]\,,
\end{align}
\end{subequations}
which generalise \cite[Eqs.\,(16)]{Qin:2020jig}.
Here, $S_g(k) =: {\mathpzc s}_A^g(k) + {\mathpzc s}_B^g(k)$,
$\{{\mathpzc s}_A^g,\gamma_5\}=0=[{\mathpzc s}_B^g,\gamma_5]$.
}

Equations~\eqref{KernelCompletion} are a pair of complex-valued integral equations whose solutions complete $K_{\rm ad}$ and hence $K^{(2)}$.
In being determined by resolving WGT identities, Eqs.\,\eqref{eq:WTI}, the results are minimal \emph{Ans\"atze} for the kernels.
This is analogous to results obtained in analyses that attempt to determine a three-point function from similar identities, \emph{e.g}., Refs.\,\cite{Ball:1980ay, Curtis:1990zs, Maris:1997tm, Bashir:2011dp, Qin:2013mta, Qin:2014vya, Aguilar:2014lha, Aguilar:2019jsj}.
There and here, notwithstanding their minimal character, the \emph{Ans\"atze} deliver substantial improvements over leading-order results; in many cases, restoring crucial symmetries, such as those leading to vector and axialvector WGT identities, that would otherwise be broken.

\begin{figure*}[!t]
\centerline{%
\includegraphics[clip, width=0.9\textwidth]{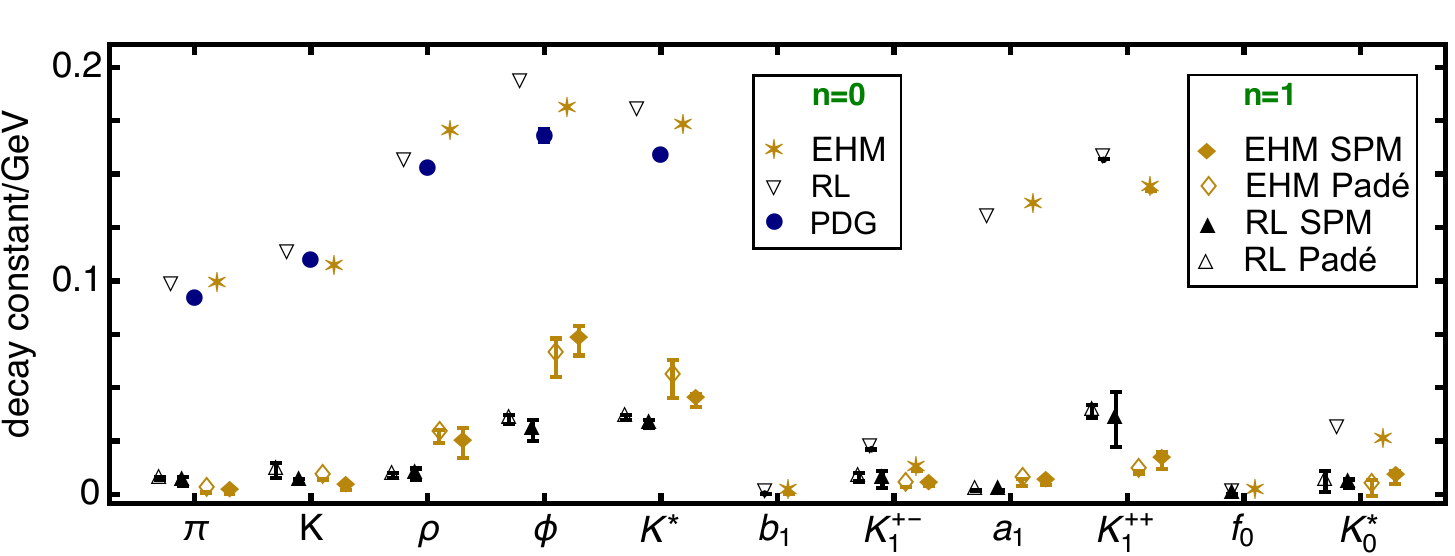}}
\caption{\label{FigLeptonic}
Leptonic decay constants for all states considered herein: ground states, $n=0$; and lowest lying radial excitations, $n=1$.
(Numerical values listed in Tables~\ref{TableDecayConstantPredictionsn0}, \ref{TableDecayConstantPredictionsn1}.)
For the excited states, we present two extrapolation results for each state, \emph{viz}.\ one obtained with Pad\'e approximants, as used for meson masses, and the other employing the Schlessinger point method \cite{Cui:2022fyr, Binosi:2018rht, Yao:2021pyf, Yao:2021pdy}.
Where available, \emph{i.e}., for some ground states, results inferred from data are also plotted \cite[PDG]{Workman:2022ynf}.
%
}
\end{figure*}

In $u$, $d$ channels, assuming isospin symmetry, \linebreak Eqs.\,\eqref{KernelCompletion} yield four real-valued functions because ${\mathsf v}_{pj}^{(\pm)}\equiv 0$.
Similarly, there are four solution functions for $s \bar s$ scattering.  In general, they are different from those associated with $u$, $d$ channels.
In $u \bar s$ and kindred channels, Eqs.\,\eqref{KernelCompletion} yield six real-valued scalar functions, the number  needed to complete the kernel in this case.
With these fourteen scalar functions in hand, then for an arbitrary vertex in the family specified by Eqs.\,\eqref{EqVertexGap}, one has symmetry-preserving EHM-improved Bethe-Salpeter \linebreak kernels for use in calculating $u$, $d$, $s$ meson bound-state properties.
It should be remembered that the solutions of Eqs.\,\eqref{KernelCompletion} depend on $P^2$; hence, must be obtained anew at each value of the total quark+antiquark momentum.  Of course, they can be obtained on a $P^2$-grid and then interpolated.

It is worth reiterating that our kernel construction can be used with any reasonable dressed-gluon-quark vertices.  Using forms more sophisticated than Eq.\,\eqref{EqVertexGap} may expand Eqs.\,\eqref{KernelCompletion} to an array of four constraint equations; but one nevertheless still arrives finally at a set of linear integral equations to solve for the coefficient functions in an expanded version of ${K}_{\rm ad}$.  Practically, therefore, little is changed.

To expose the impact of EHM-induced dressed-quark ACMs on meson properties, we solved for the spectrum of $u$, $d$, $s$ mesons as a function of $\eta$ in Eq.\,\eqref{EqVertexGap} whilst adjusting $D$ in Eq.\,\eqref{defcalG} so as to maintain $m_\rho = 0.77\,$GeV.
Given that $\eta > 0 $ adds EHM strength to the gap equation's kernel, then $D$ must decrease with increasing $\eta$ in order to achieve this outcome:
\begin{equation}
\label{RunD}
D(\eta) \stackrel{\eta \in [0, 1.6]}= D_{\rm RL} \frac{1 + 0.27 \eta }{1+1.47 \eta}\,.
\end{equation}

The spectrum obtained with $\eta = 1.2$ is displayed in Fig.\,\ref{figAllstrange}.
This value is chosen so as to simultaneously reproduce $m_{\pi,K}$ using $D_{\rm RL}$, $m_{u,s}$ specified in connection with Eq.\,\eqref{DRL}.
(The masses and amplitudes of all meson ground-states can be computed directly.  Extrapolation is used for their lightest radial excitations.)
A demonstration that the EHM-improved kernels ensure preservation of QCD's Gell-Mann--Oakes--Renner and Goldberger-Trei\-man relations \cite{GellMann:1968rz, Maris:1997tm, Qin:2014vya} may be found elsewhere \cite[Supplemental Material]{Qin:2020jig}.

Regarding the EHM-improved results in Fig.\,\ref{figAllstrange}, compared with central experimental values, the overall mean absolute relative difference is $2.9(2.7)$\%, \emph{i.e}., the EHM-improved kernels deliver a factor of $4.6$ improvement over the RL spectrum.
Further,
$m_{K^\prime} > m_{\pi^\prime}$,
$m_{\rho^\prime} > m_{\pi^\prime}$,
$m_{\rho^\prime} \approx m_{K^{\ast \prime}}$,
matching empirical results;
the $a_1$-$\rho$ and $b_1$-$\rho$ mass splittings are commensurate with empirical values because including EHM effects in the kernel has substantially increased the masses of the $b_1$ and $a_1$ mesons, whilst $m_\rho$ was deliberately kept unchanged -- Eq.\,\eqref{RunD};
$m_{\phi^\prime}-m_\phi$ matches experiment to within 2\%;
the level ordering of the $K_1^{+-}$, $K_1^{++}$ states is correct;
and quark+antiquark scalar mesons are heavy.

Given that our EHM-improved kernels deliver a realistic meson spectrum, their predictions for the masses of the as-yet unseen radial excitations of the $b_1$ and $K_1^{++}$ states should be reasonable (in GeV):
\begin{subequations}
\label{newmasses}
\begin{align}
m_{b_1^\prime} & = 1.67(3)\,, \;
m_{b_1^\prime} -m_{b_1} = 0.51(3)\,, \\
m_{K_1^{++ \prime}} & = 1.63(1)\,,\;
m_{K_1^{++ \prime}} - m_{K_1^{++}} = 0.33(1) \,.
\end{align}
\end{subequations}
The mass splittings in the partner channels are:
$m_{a_1^\prime} -m_{a_1} = 0.47(4)$,
$m_{K_1^{+- \prime}} - m_{K_1^{+-}} = 0.39(3)$.
Another potentially useful observation is that the EHM-improved kernels predict $m_{b_1^\prime} \approx m_{a_1^\prime}$, $m_{K_1^{++\prime}}\approx m_{K_1^{+-\prime}}$.

Employing the Bethe-Salpeter equation solutions \linebreak used to produce the spectrum, with the amplitudes canonically normalised in the standard fashion \linebreak \cite[Sec.\,3]{Nakanishi:1969ph}, one can calculate the leptonic decay constant, $f_H$, associated with each state.
The definitions of these observables may be found, \emph{e.g}., in Refs.\,\cite[Eqs.\,(2)]{Yin:2021uom}, \cite[Eq.\,(6)]{Bhagwat:2006py}.  They reveal that such decay constants are identically zero for $J^{PC}=0^{++}$, $1^{+-}$ states constituted from mass-degenerate valence degrees-of-freedom.
Regarding radially excited states, we choose the phase by requiring that the zeroth Chebyshev moment of the leading term in the system's Bethe-Salpeter amplitude is positive in the ultraviolet.  This is opposite to the convention in Ref.\,\cite{Holl:2005vu}.
Our results are collected in Fig.\,\ref{FigLeptonic}, drawn for easy assimilation, and detailed in Tables~\ref{TableDecayConstantPredictionsn0}, \ref{TableDecayConstantPredictionsn1}.

\begin{table}[!t]
\caption{\label{TableDecayConstantPredictionsn0}
Ground-state  ($n=0$) leptonic decay constants (in GeV) calculated using RL and EHM-improved kernels compared with values derived from Ref.\,\cite[PDG]{Workman:2022ynf}, where available.
Decay constants of $1^{+-}$ and $0^{++}$ states are identically zero for systems built from mass-degenerate valence degrees-of-freedom.
%
%
 }
\begin{center}
\begin{tabular*}
{\hsize}
{
l@{\extracolsep{0ptplus1fil}}
|l@{\extracolsep{0ptplus1fil}}
l@{\extracolsep{0ptplus1fil}}
l@{\extracolsep{0ptplus1fil}}}
meson~$n=0$\;  &RL & PDG & EHM-improved \\\hline
$\pi$ & $0.097\ $ & $0.092(1)\ $ & $0.097\ $\\
 $K$ & $0.112\ $ & $0.110(1)\ $ & $0.105\ $\\
$\rho$ & $0.155\ $ & $0.153(1)\ $ & $0.168\ $\\
$\phi$ & $0.192\ $ & $0.168(3)\ $ & $0.179\ $\\
$K^\ast$ & $0.179\ $ & $0.159(1)\ $ & $0.171\ $\\
$b_1$ &$0\ $ & & $0\ $\\
$K_1^{+-}$ & $0.021\ $ & & $0.011\ $\\
$a_1$ & $0.129\ $ & & $0.134\ $\\
$K_1^{++}$ & $0.157\ $ & & $0.142\ $\\
$f_0$ & $0\ $ & & $0\ $\\
$K_0^\ast$ & $0.03\ $ & & $0.024\ $\\\hline
\end{tabular*}
\end{center}
\end{table}

For ground states, the leptonic decay constants can be calculated directly.  However, as with meson masses, the decay constants for excited states must be obtained by extrapolation.  To achieve that we wrote $f_H(\lambda(m))$, where $\lambda(m)$ is the eigenvalue used to define the Bethe-Salpeter equation, computed $f_H(\lambda(m))$ on a set of equally distributed $m$-values within the domain on which the Bethe-Salpeter amplitude can be straightforwardly obtained, then extrapolated the result to $\lambda(m) = 1$.

Regarding the extrapolation, we provide two sets of results.  One was obtained with the Pad\'e-approximant method used above to determine meson masses.  The other set was calculated using the Schlessinger point method (SPM), discussed in detail elsewhere \cite{Cui:2022fyr, Binosi:2018rht, Yao:2021pyf, Yao:2021pdy}.  Herein, to calculate the leptonic decay constants of $n=1$ mesons, we worked with $N=40$ values of $f_H(\lambda(m))$ and $M=20$-element subsets to make available a collection of $\approx 138$-billion continued-fraction interpolants, from which we randomly selected 10\,000 functions that were smooth and single-sign on the domain of interpolation and extrapolation.  Extrapolating each of these functions to $\lambda(m)=1$ yields the quoted mean value and uncertainty, which is the standard deviation of that mean.
(The terminology here matches that in Ref.\,\cite{Cui:2022fyr}.)
Owing to its foundations in analytic function theory and the powerful statistical aspect introduced in modern applications, the SPM provides reliable extrapolations with a rigorously quantified uncertainty.  The important observation here is that both methods yield consistent results in all cases.

Given that we have simplified the interaction in Eq.\,\eqref{defcalG}, keeping only the first term, then the comparison between our results for ground-state meson decay constants and the few known empirical values is favourable, especially since decay constants are sensitive to ultraviolet physics, which we have omitted.  Further, there is a hint that the EHM-improved kernels deliver better agreement.

The results for radially excited states are especially interesting.  In quantum mechanics models of positro\-nium-like systems, one normally finds that, owing to zeros in the associated radial wave functions, the decay constant of a first radial excitation is $(1/8)$-times that of the ground state.  Our Bethe-Salpeter equation predictions are broadly consistent with this pattern, except for $J^P=0^-$  mesons.  In these channels, the leptonic decay constant must vanish in the chiral limit \cite{Holl:2005vu, McNeile:2006qy, Ballon-Bayona:2014oma}.  Whilst our results are in accord with this prediction, quantum mechanics models cannot deliver such an outcome; thus, our values for unmeasured decay constants warrant testing.

\begin{table}[!t]
\caption{\label{TableDecayConstantPredictionsn1}
Leptonic decay constants (in GeV) of radial excitations ($n=1$) calculated using RL and EHM-improved kernels.  Empirical data are lacking.  Here we present two extrapolation results for each state, \emph{viz}.\ one obtained with Pad\'e approximants, as used for masses, and the other employing the Schlessinger point method \cite{Cui:2022fyr, Binosi:2018rht, Yao:2021pyf, Yao:2021pdy}.
For systems constituted from mass-degenerate valence degrees-of-freedom, the decay constants of $1^{+-}$ and $0^{++}$ states are identically zero.
(Calculated values show the relevant extrapolation uncertainty.)
 }
\begin{center}
\begin{tabular*}
{\hsize}
{
l@{\extracolsep{0ptplus1fil}}
|l@{\extracolsep{0ptplus1fil}}
l@{\extracolsep{0ptplus1fil}}
|l@{\extracolsep{0ptplus1fil}}
l@{\extracolsep{0ptplus1fil}}}
meson\; & \multicolumn{2}{c|}{RL} & \multicolumn{2}{c}{EHM-improved} \\
$n=1\ $ & Pad\'e & SPM & Pad\'e & SPM \\\hline
$\pi$ & $0.0073(06)\ $ & $0.006(2)\ $ & $0.0008(3)\ $ & $0.0010(06)\ $ \\
$K$ & $0.0112(34)\ $ & $0.0061(11)\ $ & $0.0071(2)\ $ & $0.0033(14)\ $ \\
$\rho$ & $0.0088(11)\ $ & $0.0095(26)\ $ & $0.027(3)\ $ & $0.024(7)\ $ \\
$\phi$ & $0.035(2)\ $ & $0.030(5)\ $ & $0.064(9)\ $ & $0.072(7)\ $ \\
$K^\ast$ & $0.036(1)\ $ & $0.033(2)\ $ & $0.054(9)\ $ & $0.044(3)\ $ \\
$b_1$ & $0\ $ & $0\ $ & $0\ $ & $0\ $\\
$K_1^{+-}$ & $0.008(2)\ $ & $0.007(4)\ $ & $0.0034(1)\ $ & $0.0042(1)\ $ \\
$a_1$ & $0.0018(2)\ $ & $0.0017(2)\ $  & $0.0056(17)\ $ & $0.0056(12)\ $ \\
$K_1^{++}$  & $0.039(3)\ $ & $0.035(13)\ $& $0.0101(7)\ $ & $0.016(4)\ $ \\
$f_0$ & $0\ $ & $0\  $ & $0\ $ & $0\ $ \\
$K_0^\ast$ &$0.006(5)\ $ & $0.005(2)\ $& $0.0030(36)\ $ & $0.008(3)\ $ \\\hline
\end{tabular*}
\end{center}
\end{table}

\section{Perspectives}
\label{epilogue}
It is worth reiterating that the method employed herein to calculate Bethe-Salpeter kernels for meson bound-state problems is both flexible and certain to give results that are symmetry consistent with any realistic gluon-quark vertex, $\Gamma_\nu$.  This is true whether or not the diagrammatic content of $\Gamma_\nu$ is known.  Hence, the approach paves a way to new synergies between continuum and lattice approaches to strong interactions.

The kernels are not unique; but they are closed-form \emph{Ans\"atze} that enable one to reliably reveal and understand how key features of emergent hadron mass (EHM), contained in $\Gamma_\nu$, are expressed in the meson spectrum.  As an example, we highlighted the multifarious impacts on meson properties of the simple fact that EHM forces dressed-quarks to posses a large anomalous chromomagnetic moment.  Extending these ideas to baryon bound-state problems would be valuable and attempts are underway.

This study delivers the first Poincar\'e-invariant treatment of the spectrum and decay constants of the ground- and first-excited states of $u$, $d$, $s$ mesons along with predictions for masses of as-yet unseen states and many unmeasured decay constants.  To expedite the analysis, we used a simplified treatment of quark-antiquark scattering.  It would therefore be natural to repeat the work using a more realistic interaction.  Also, having benchmarked the method against known lighter-quark states, extension of the approach to heavy+light mesons \cite{Binosi:2018rht, Chen:2019otg, Qin:2019oar}, hybrid mesons \cite{Burden:2002ps, Qin:2011xq, Hilger:2015hka, Xu:2018cor} and glueballs \cite{Meyers:2012ka, Souza:2019ylx, Kaptari:2020qlt, Huber:2021yfy} is  desirable, especially given world-wide investments in studies of and searches for such states at high-energy, high-luminosity facilities \cite{BESIII:2020nme, Anderle:2021wcy, AbdulKhalek:2021gbh, Pauli:2021gde, Quintans:2022utc}.

\medskip
\noindent\textbf{Acknowledgments}.
We are grateful for constructive comments from \mbox{C.~Xu}.
Work supported by:
National Natural Science Foundation of China (grant nos. \linebreak 12135007, 11805024, 12233002);
and
Natural Science Foundation of Jiangsu Province (grant no.\ BK20220122).
%
%


\medskip
\noindent\textbf{Data Availability Statement}.
This manuscript has no associated data or the data will not be deposited. [Authors’ comment: All information necessary to reproduce the results described herein is contained in the material presented above.]

\appendix

\section{Supplementary Equations}
In passing from Eqs.\,\eqref{eq:WTI} to \eqref{KernelCompletion}, we follow the route described in Ref.\,\cite[Supplemental Material]{Qin:2020jig}.  The key results are \cite[Eqs.\,(S.6)]{Qin:2020jig}, whose generalisations to $m_g\neq m_h$ are recorded here:
\begin{subequations}
\label{SuppEqs}
\begin{align}
\Sigma_{B}^g&\left(k_{+}\right)  = \int_{d q}\left\{-K_{L 0}^{(+)} \left[\Delta_{\mathpzc{s}_A^{gh}}^{\pm}(q)\right]K_{R 0}^{(-)} \right. \nonumber \\
& \left. -K_{L 1}^{(-)} \left[{\mathpzc{s}}_{B}^g\left(q_{+}\right)\right]  K_{R 1}^{(-)}
+K_{L 1}^{(+)} \left[{\mathpzc{s}}_{B}^h\left(q_{-}\right)\right]  K_{R 1}^{(+)}\right\},
\label{S.6a} \\
0& =\int_{d q}\left\{K_{L 0}^{(+)} \left[{\mathpzc{s}}_{B}^h\left(q_{-}\right)\right] K_{R 0}^{(-)}\right.  \nonumber \\
& \left. -K_{L 0}^{(-)} \left[{\mathpzc{s}}_{B}^g\left(q_{+}\right)\right] K_{R 0}^{(+)}
-K_{L 2}^{(+)} \left[\Delta_{\mathpzc{s}_A^{gh}}^{\pm}(q)\right] K_{R 2}^{(+)}\right\},
\label{S.6b} \\
%
\Delta_{\Sigma_A^{gh}}^\pm &
= \int_{d q}\left\{ -K_{L 0}^{(+)} \left[{\mathpzc{s}}_{B}^g\left(q_{+}\right)\right] K_{R 0}^{(-)} \right. \nonumber \\
& \left. +K_{L 0}^{(-)} \left[{\mathpzc{s}}_{B}^h\left(q_{-}\right)\right] K_{R 0}^{(+)}  - K_{L 2}^{(-)} \left[\Delta_{\mathpzc{s}_A^{gh}}^{\pm}(q)\right] K_{R 2}^{(-)}\right\},
\label{S.6c} \\
-\Sigma_{B}^h&\left(k_{-}\right)  =\int_{d q}\left\{-K_{L 0}^{(-)} \left[\Delta_{\mathpzc{s}_A^{gh}}^{\pm}(q)\right] K_{R 0}^{(+)} \right. \nonumber\\
& \left. + K_{L 1}^{(-)} \left[{\mathpzc{s}}_{B}^h\left(q_{-}\right)\right] K_{R 1}^{(-)} - K_{L 1}^{(+)} \left[{\mathpzc{s}}_{B}^g\left(q_{+}\right)\right] K_{R 1}^{(+)}
\right\},
\label{S.6d}
\end{align}
\end{subequations}
where $S_g(k) =: {\mathpzc s}_A^g(k) + {\mathpzc s}_B^g(k)$,
$\{{\mathpzc s}_A^g,\gamma_5\}=0=[{\mathpzc s}_B^g,\gamma_5]$.
In writing these identities, the Bethe-Salpeter kernel has been decomposed into a direct-product form, also grouped according to $\gamma_5$ commutation properties:
\begin{eqnarray}
	{K}^{(2)} &=:& \left[ K_{L0}^{(+)} \otimes K_{R0}^{(-)} \right] + \left[ K_{L0}^{(-)} \otimes K_{R0}^{(+)} \right] \notag \\
	&& + \left[ K_{L1}^{(-)} \otimes_+ K_{R1}^{(-)} \right] + \left[ K_{L1}^{(+)} \otimes_+ K_{R1}^{(+)} \right] \notag \\
	&& + \left[ K_{L2}^{(-)} \otimes_- K_{R2}^{(-)} \right] + \left[ K_{L2}^{(+)} \otimes_- K_{R2}^{(+)} \right]\,,
	\label{eq:kernel5}
\end{eqnarray}
where $\otimes_\pm := \tfrac{1}{2}(\otimes \pm \gamma_5 \otimes \gamma_5)$ and $\gamma_5 X^{(\pm)}\gamma_5 = \pm X^{(\pm)}$ (see Ref.\,\cite[Eqs.\,(3), (9)]{Qin:2020jig} for additional guidance).






%


\end{document}